\documentclass[10pt]{article}
\usepackage{graphics}
\usepackage{amsmath}
\usepackage{amssymb}

\newcommand{\ie}{\textit{i.e.}~}
\newcommand{\eg}{\textit{e.g.}~}

\newcommand{\hpsi}{\hat{\Psi}}
\newcommand{\hchi}{\hat{\chi}}

\newcommand{\kkk}{\mathrm{k}}
\newcommand{\xxx}{\mathrm{x}}

\newcommand{\yyy}{\mathrm{y}}

\newcommand{\mmm}{\mathfrak{m}}
\newcommand{\nnn}{\mathfrak{n}}

\title{Emergent gravitational dynamics in Bose--Einstein condensates\footnote{To appear in the proceedings of the XXV Max Born Symposium, 28/6/2009-3/7/2009, Wroclaw, Poland}}

\author{Lorenzo Sindoni$^{1}$\footnote{e-mail: \texttt{sindoni@sissa.it}}, Florian Girelli$^{2}$\footnote{e-mail: \texttt{girelli@physics.usyd.edu.au}}, Stefano Liberati$^{1}$\footnote{e-mail: \texttt{liberati@sissa.it}}\\
   $^{1}$ \textit{SISSA and INFN, Trieste}, \\
   $^{2}$ \textit{School of Physics, Sydney University}}

\begin{document}

\maketitle
\begin{abstract}
We discuss a toy model for an emergent non-relativistic gravitational theory. Within a certain class of Bose--Einstein condensates, it is possible to show that, in a suitable regime, a modified version of non-relativistic Newtonian gravity does effectively describes the low energy dynamics of the coupled system condensate/quasi-particles.
\end{abstract}


\section{Introduction}

In recent years the emergent gravity approach has constantly gained momentum \cite{HuMicro}. In this particular perspective, the gravitational field, encoded in the spacetime geometry and its dynamics, is seen as a kind of large number/thermodynamical limit of some more fundamental theory.

While there is no proof that this is the case, there are  evidences to support  this point of view. First of all, let us remind the striking correspondence existing between the laws of black hole mechanics in general relativity and the laws of thermodynamics \cite{BardeenCarterHawking}. This correspondence has inspired a deeper study of the thermodynamical aspects of gravitation. In particular, it has been shown that Einstein's equation can be seen as some sort of equation of state of a thermodynamical system at equilibrium \cite{Jacobson,ElingGuedensJacobson}. Furthermore, from the study of the properties of gravitational collapse and the resulting black hole formation, very specific patterns have been recognized  in the parameter space of the initial conditions which resemble the behavior of critical phenomena (phase transitions) \cite{Choptuik,CriticalCollapseLLR}.

In addition, it has been realized that within several condensed matter systems it is possible to distinguish some peculiar regimes in which the effective degrees of freedom are represented by fields propagating over effective pseudo-Riemannian structures \cite{AnalogueReview}.

For instance, in the case of perfect, irrotational and barotropic fluids, it can be proved that the perturbations in the velocity potential (\ie the scalar function $\theta$ whose gradient gives the velocity of the fluid, $\vec{v} \propto \vec{\nabla} \theta$) do obey a massless Klein--Gordon equation in a curved effective spacetime whose metric tensor is given by the so-called acoustic metric,
\begin{equation} \label{eq:acoustic}
g_{\mu\nu} = \frac{\rho}{c_{s}^{2}} \left( \begin{array}{ccc}
-(c_{s}^{2} - v^{2}) & \vdots & v^{i} \\
\cdots & \cdot & \cdots \\
v^{i} & \vdots& \delta_{ij}
\end{array}
\right),
\end{equation}
where $\rho$ is the local density of the fluid, $c_{s}$ it the (local) speed of sound and $v_{i}$ is the velocity field of the fluid flow.

Analogue models have been used to understand (and possibly to test in a laboratory) some peculiar aspects of physics in curved spacetimes, otherwise inaccessible (\eg Hawking radiation). For a review of the subject see \cite{AnalogueReview}. For the large majority, these analogue models for gravity do offer the possibility of studying some kinematical aspects of physics of curved spacetimes, leaving aside the issue of dynamics.

Despite the fact that, to date, there are no completely successful models of emergent gravity, \ie non-gravitational system possessing an effective General Relativity (GR) like limit, it is interesting to study simple toy models in which some aspects of the gravitational interaction can be mimicked. Here we present a toy model \cite{BEC}, based on a well-known condensed matter system, a Bose--Einstein condensate of dilute, weakly interacting gas of bosons \cite{blackbook}. While this toy model is very far from representing an analogue of a realistic theory of gravity, it gives nonetheless some insights on some interesting features of emergent theories.

The behaviour of dilute, weakly interacting Bose gases can be described with the formalism of second quantization, by introducing the field operators
describing the gas atoms {($V$ is the volume of the box in which we consider the gas)} \begin{equation}
\hpsi(\xxx) = \frac{1}{\sqrt{V}} \sum_{\kkk} \hat{a}_{\kkk} e^{i\kkk\cdot \xxx},
\end{equation}
and  a suitable Hamiltonian:
\begin{equation}\label{Hamiltonian0}
\hat{H}_{0} = \int d^{3} x \, \hpsi^{\dagger}(\xxx) \left(  \frac{-\hbar^{2} \nabla^{2}}{2m} - \mu + \frac{\kappa}{2} |\hpsi(\xxx)|^{2}  \right)\hpsi(\xxx).
\end{equation}
As it is well known, for $\mu,\kappa$ positive, this Hamiltonian leads to a condensation, \ie the ground state of the system is characterized by a macroscopic occupation number of a single particle state. Within the formalism, this corresponds to the field operator $\hpsi$ having a non-vanishing vacuum expectation value,
\begin{equation}
\hpsi(\xxx)\approx \psi(\xxx) \mathbb{I} + \hchi(\xxx), \qquad \langle \Omega | \hpsi(\xxx) | \Omega \rangle = \psi(\xxx),
\end{equation}
where $\psi$ is a classical complex scalar field, the condensate wavefunction, describing the mean field, and $\hchi$ is a quantum operator describing the residual quantum fluctuations around the condensate, or, in more physical terms, the atoms which are out of the condensate.

Obviously, the nonlinearity in the Hamiltonian \eqref{Hamiltonian0} makes the analysis of the physical properties of this system rather involved. As a first approximation, it is useful to assume that the non-condensed fraction is small compared to the condensed fraction. Consequently, one can introduce the parametrization
\begin{equation}
\hpsi = \psi + \epsilon \hchi + O(\epsilon^{2}),
\end{equation}
where $\epsilon \ll 1$ is a dimensionless parameter which is introduced for bookkeeping, and organize an expansion in powers of $\epsilon$ of the equation of motion.
\paragraph{Gross--Pitaevskii}
The lowest order equation describes the condensate wave-function alone, without the backreaction terms induced by the $\hchi$ field. It is called the Gross--Pitaevski equation and it is giving the first approximation to the condensate wavefunction dynamics. It is obtained from the equations of motion for the system by the replacement $\hpsi \rightarrow \psi$. In the setup we are considering, the equation reads:
\begin{equation}\label{GP}
i\hbar\frac{\partial \psi}{\partial t} = -\frac{\hbar^{2}}{2m }\nabla^{2} \psi - \mu \psi + \kappa |\psi|^{2} \psi,
\end{equation}
which admits constant solutions, $\psi=(\kappa/\mu)^{1/2}$, describing homogeneous condensates. This kind of solutions, in the analogue model perspective, correspond to a flat acoustic metric, \ie to Minkowski spacetime, and therefore will play a particular role in the following. The weak field limit of GR will correspond to the weak field limit around this constant configuration. In fact, as a direct calculation  show, the correspondence between the acoustic metric $g_{\mu\nu}$ and the  complex scalar field $\psi$ is such that:
\begin{equation}
\left.\begin{array}{c}g_{\mu\nu}\approx \eta_{\mu\nu} + h_{\mu\nu}\\
\psi \approx (\kappa/\mu)^{1/2}(1+u(x)+iv(x))
\end{array}\right\} \Rightarrow h_{00} \varpropto u(x)
\end{equation}
so that the perturbations in the number density encoded in $u(x)$ are the natural candidate to represent some sort of Newtonian gravitational field.

\paragraph{Quasiparticles}
The next order equation is an equation for the noncondensed fraction $\hchi$ in the external field $\psi$. The equation of motion has the shape:
\begin{equation} \label{oldparticles}
i\hbar\frac{\partial}{\partial t} \hchi = -\frac{\hbar^{2}}{2m } \nabla^{2} \hchi + \mu \hchi + \mu \hchi^{\dagger}.
\end{equation}
The Hamiltonian \eqref{Hamiltonian0} does possess a global $U(1)$ symmetry, $\hpsi \rightarrow e^{i\alpha} \hpsi$, which is spontaneously broken by the
 condensation mechanism (leading to the $\psi \neq 0$ ground state). As a consequence of the Goldstone theorem, the excitations will be gapless. (This 
 can also be proved directly by diagonalizing the equation \eqref{oldparticles} using the Bogoliubov transformations.) 
Finally, it can be showed that, in some low energy regime, quasiparticles are described by a massless scalar field propagating on an effective acoustic geometry of the same shape of \eqref{eq:acoustic}.

\section{The coupled system}
In order to see how some sort of gravitational dynamics is encoded in the BEC, a suitable dynamical framework must be set up in order to see how the
quasiparticles backreact over the condensate. This formalism consists in considering an improved version of the Gross--Pitaevski equation which consistently takes into account the effect of the particles out of the condensate.
The Gross--Pitaevski (GP) equation is replaced by the so-called Bogoliubov--de Gennes (BdG) equation
\begin{equation}\label{BdG}
i\hbar\frac{\partial }{\partial t} \psi = -\frac{\hbar^{2}}{2m }\nabla^{2} \psi - \mu \psi + \kappa |\psi|^{2} \psi + 2 \kappa \nnn \psi + \kappa \mmm \psi^{*},
\end{equation}
where $\nnn,\mmm$ are given by the expectation values:
\begin{equation}\label{eq:sources}
\nnn = \langle \Xi  |\hchi(x)^{\dagger} \hchi(x)| \Xi \rangle, \qquad \mmm = \langle \Xi|\hchi(x)^{2}| \Xi\rangle,
\end{equation}
where the state $|\Xi \rangle$ is the particular state one is considering. Notice that, if this state were the Fock vacuum state for particles, these expectation values would be identically zero. Notice also that one is implicitly taking a normal ordering in the particle operator, so that an unphysical (divergent) zero point energy is removed automatically.

This equation, suitably modified, will give rise to the dynamics for the field $u(x)$, and hence to an analogue of the Poisson equation.

To construct some analogue of Newtonian gravity, we need massive particles as sources of the gravitational field (massless particles do not gravitate in Newtonian gravity). Therefore, the quasiparticles must not be Goldstone bosons, but instead pseudo-Goldstones: the $U(1)$ symmetry has to be broken explicitly at the level of the Hamiltonian. In order to do this, we add an extra term  to the Hamiltonian,
\begin{equation}
\hat{H}_{0}\rightarrow \hat{H} = \hat{H}_{0} + \hat{H}_{\lambda}, \qquad \hat{H}_{\lambda} =- \frac{\lambda}{2} \int d^{3}x \left( \hpsi(x)^{2}+(\hpsi^{\dagger}(x))^{2} \right),
\end{equation}
where $\lambda$ is a coupling constant having the dimensions of an energy in these choice of units. The breaking of the $U(1)$ symmetry has an obvious interpretation: the number of bosons is no longer a conserved charge. This could lead to some issues about the physical realizability of such a system. For a critical discussion on this point see \cite{BEC}.

Preliminary analysis of the condensate wavefunction based on the extension of the GP equation to this case, leads to the homogeneous solution
\begin{equation}
\psi= \sqrt{n_{c}} e^{-i\theta/\hbar},\qquad n_{c} = \frac{\mu+\lambda}{2}, \qquad \theta = 0,
\end{equation}
where the phase $\theta=0$ is fixed by stability of the condensate itself: for different values of $\theta$ the quasiparticles would show a tachyonic instability.

The analysis of the properties of the quasiparticles in the case of homogeneous background (see \cite{BEC} for details) leads to the conclusion that the quasiparticles dispersion relation is
\begin{equation}
E= \left(\mathcal{M}^{2}c_{s}^{4} + p^{2}c_{s}^{2} + \frac{p^{4}}{4m^{2}}\right)^{1/2},
\end{equation}
where
\begin{equation}
c_{s}^{2} = \frac{\mu+2\lambda}{m}, \qquad \mathcal{M}^{2} = 4\frac{\lambda(\mu+\lambda)}{(\mu+2\lambda)^{2}} m^{2}.
\end{equation}
In the case of small momenta, and when the condensate wavefunction is not exactly homogeneous (\ie when $u(\xxx) \neq 0$), the Hamiltonian for the quasiparticles takes the shape
\begin{equation}
\hat{H}_{q.p.} \approx \mathcal{M}c^{2}_{s} -  \frac{\hbar^{2}\nabla^{2}}{2\mathcal{M}}+ 2\frac{(\mu+\lambda)(\mu+4\lambda)}{\mathcal{M}c_{s}^{2}}u(\xxx),
\end{equation}
which leads to the identification of a  ``gravitational potential'':
\begin{equation}
\Phi_{grav}(\xxx) = \frac{(\mu+4\lambda)(\mu+2\lambda)}{2\lambda m} u(\xxx).
\end{equation}

\subsubsection{The emergent gravitational system}

Having presented the main ideas and required tools, we pass to the results. Consider the Hamiltonian with the $U(1)$ breaking term. In the limit in which the backreaction of the condensate is small, \ie in the limit in which there are few quasiparticles, when the condensate is almost homogeneous, the Bogoliubov--de Gennes equation can be rewritten as:

\begin{equation}
\left( \nabla^{2} - \frac{1}{L^{2}} \right) \Phi_{grav} = 4 \pi G_{N} \rho_{\mathrm{matter}} + \Lambda,
\end{equation}
where
\begin{eqnarray}
&& G_{N} \equiv \frac{\kappa(\mu+4\lambda)(\mu+2\lambda)^{2}}{4 \pi \hbar^{2} m \lambda^{3/2} (\mu+\lambda)^{1/2}} ,\qquad
 \Lambda \equiv \frac{2 \kappa(\mu+4\lambda)(\mu+2\lambda)}{\hbar^{2} \lambda}\left(\nnn_\Omega +\frac{1}{2} \mmm_\Omega\right), \label{eq:lambda}\\
&& \label{eq:range}
L^{2} \equiv \frac{\hbar^{2}}{4m(\mu+\lambda)}.
\end{eqnarray}
Notice the peculiar splitting of the source term. A detailed analysis \cite{BEC} shows that the expectation values \eqref{eq:sources}
always split into two contributions, one nonlocal term due to the  quasiparticles  ($\rho_{\mathrm{matter}}$), and an unavoidable vacuum contribution, $\Lambda$, due to the inequivalence between the Fock vacuum for particles and the Fock vacuum for quasiparticles.

The reader will immediately realize that this interaction is very far away from a realistic Newtonian gravitational interaction: indeed,  the would-be Poisson equation includes a term which makes the interaction short ranged. In particular, this range is set by the healing length $L$, which is an UV scale for the physics of the BEC (very much like the Planck scale in quantum gravity). This might have been guessed from the beginning, since that the healing length represents the typical scale for the dynamics of the condensate. Despite the fact that this system is not an analogue for a realistic form of gravitational interaction, it does offer some intriguing inspirations which we develop now.

\section{Quasi-particles and locality}

An aspect of the physics of quasiparticles which is not often stressed concerns the issue of locality, \ie whether the effective Lagrangian of the quasiparticles does obey the axioms of local quantum field theory.

In BEC, the particles and quasiparticles field operators are related by a Bogoliubov transformation, which have the following general structure:
\begin{equation}
\omega^{A}(\kkk) = M^{A}_{B}(\kkk) \phi^{B}(\kkk),
\end{equation}
when working in momentum space. This relation is always linear. However, it is nonlocal:
\begin{equation}\label{eq:realspacebogolubov}
\omega^{A}(\xxx) = \int d^{3}y\, K^{A}_{B}(\xxx,\yyy) \phi^{B}(\yyy),
\end{equation}
where the kernel $K$ is determined by the Bogoliubov coefficients:
\begin{equation}
K^{A}_{B}(\xxx,\yyy) = \int d^{3}k \, M^{A}_{B}(\kkk) e^{-i\kkk \cdot (\xxx-\yyy)}.
\end{equation}
This is the mathematical statement of the fact that quasi-particles are collective degrees of freedom. The structure of the transformation immediately implies that there is a mismatch between the notion of locality of the quasi-particle with respect to the notion of locality of the atoms.
As one easily realizes, the two classes of operators $\phi,\omega$, separately, do obey canonical equal time commutation relations:
\begin{equation}
[\phi^{A}(\xxx),(\phi^{B})^{\dagger}(\yyy)] = \delta^{AB} \delta^{3}(\xxx-\yyy), \qquad [\omega^{A}(\xxx),(\omega^{B})^{\dagger}(\yyy)] = \delta^{AB} \delta^{3}(\xxx-\yyy),
\end{equation}
which are a direct consequence of the fact that Bogoliubov transformations are preserving the algebra of the creation-annihilation operators.
Therefore, as long as we use only one of the two families, there is no way in which a deviation from standard local quantum field theory can be manifest.
However, the mixed commutators are nontrivial. It is straightforward to see that:
\begin{equation}
[\phi^{A}(\xxx), (\omega^{B})^{\dagger}(\yyy)] = (K^{AB})^{*}(\xxx,\yyy),
\end{equation}
Of course, this fact becomes crucial when the effective Lagrangian describing the physics of the quasi-particles involves terms mixing particle and quasi-particle operators.

\begin{figure}
\hspace{1.5cm}
\includegraphics{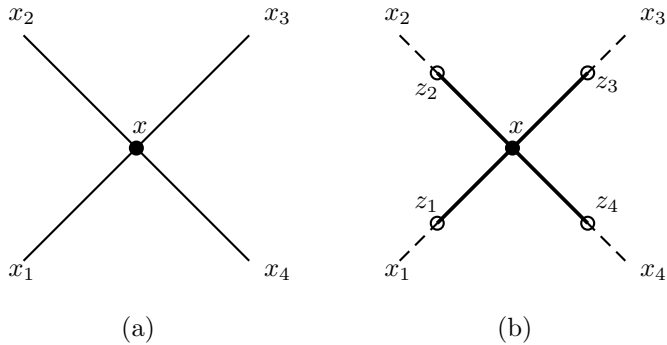}
\caption{(a) Feynman diagram of a four-particles vertex. (b) Lowest order Feynman diagram for the interaction of four quasi-particles (dashed lines),
induced by the interactions of particles (solid line). The circles denote the conversion of particles into quasi-particles at intermediate points. They represent the action of the Bogoliubov transformations.}\label{fig:feynman}
\end{figure}

In the case of the BEC it is pretty clear how the underlying dynamics induces in the action for the quasi-particles an interaction term of the form
\begin{equation}
L_{\mathrm{int}} = -\frac{\kappa}{4} \left((\phi^{A})^{\dagger}\phi_{A} \right)^{2}
\end{equation}
which explicitly involves the particle fields, rather than the quasi-particles. Hence, when computing the effects of the interaction terms, \eg scattering processes between the quasi-particles, the nonlocality encoded in the kernel $K$ will necessarily enter the physical quantities (see figure \ref{fig:feynman}).

The (effective) quantum field theories describing high energy physics are based on the assumption on locality, which is entering the physical observables at different levels.
A detection of a violation of locality could be an indication that the (effective) quantum field theory we are using at low energies, namely the standard model of particle physics, is an effective field theory of quasi-particles over a condensate. Since we know that it is impossible to obtain the entire condensate dynamics out of a local quantum field theory based on quasi-particles (without knowing the Bogoliubov transformations), we could imagine that if the same picture holds for the real world, and in particular for gravitons, the difficulties in quantizing gravity are due to the fact that we are insisting using quasi-particle basis instead of using the particle one.

It is interesting to note that the nonlocality is controlled by a very specific energy scale. We have seen that the Bogoliubov coefficients depend on the momentum of the quasi-particles in such a way that quasi-particles operators are becoming particle operators for very high momenta. The scale of the crossover is given by the healing length scale, which represent then the typical scale associated with nonlocality.

\section{Conclusions}

The model presented, obviously, is not a model of emergent gravity in the sense of GR. There are important conceptual and practical obstructions which forbid the BEC system to be used to discuss analogue of the gravitational field. In this sense, it is an attempt doomed to fail from the very beginning. First of all, the model is non-relativistic. Despite some encouraging results towards the extension to a relativistic theory made in \cite{timenordstrom}, it is still not clear how to evade the Weinberg--Witten theorem \cite{WW,Loebbert,Jenkins2} in order to produce a viable model of emergent spin-2 gravity. Nevertheless, the brief discussion of the BEC-based toy model offers the possibility of showing very nice properties which could be of help in understanding some of the puzzles we find in the study of the gravitational field.

The two main lessons which we should take concern the fate of locality and the smallness of the cosmological constant. As far as locality is concerned, there are two levels in a BEC:  the particles together with their local interactions, and  the quasiparticles which are the effective degrees of freedom propagating in the condensed phase. These latter are some collective degrees of freedom obtained from the non-local Bogoliubov transformations: the nonlocality of the relationship between particles and quasiparticles is translated into the failure of the interactions between quasiparticles to be described by a local quantum field theory.

Furthermore, when the quasiparticles are seen as sources of the ``gravitational field'' encoded into the condensate wavefunction, the nonlocality manifests itself again in the non-local way in which they are coupled to it.

A second important aspect which could be of interest in the discussion of the cosmological constant problem is the naturalness of the smallness of the cosmological constant term present in a BEC. As we have seen, the cosmological constant term is proportional to the depletion factor, \ie by the ratio between the number of particles on excited states and the number of particles in the ground state (the condensate). Whenever this ratio is small, the cosmological constant is similarly suppressed.

Notice that the smallness of the depletion factor depends on two properties: the dilution of the bosons and the smallness of the scattering lengths.

The relation between the analogue of the cosmological constant and the depletion factor is intriguing. In fact, the smallness of the cosmological constant term is intrinsically related to the quality of the condensate phase, \ie the larger is the condensation the smaller is the depletion factor. Accordingly, the smaller is the cosmological constant, the better is the mean field approximation, \ie the way in which the condensate wavefunction encodes the dynamics of the condensate. Since the mean field is tied to the acoustic metric, the smallness of the ``cosmological constant'' term is directly related to how well the acoustic geometry will be able to describe the dynamics of the quasiparticles.

This discussion should reinforce the idea that the emergent gravity program, and in particular the analysis of even simple toy models, could shed a light on very deep aspects of physics of spacetime beyond the grasp of standard approaches like general relativity.

\begin{center}{\bf Acknowledgments}\end{center}
L.S. wants to thank the organizers of the conference for their kind hospitality.

\bibliographystyle{aipprocl}

\end{document}